\documentstyle[11pt,paspconf,epsf,astrobib]{article}

\makeatletter
\def\thebibliography{\section*{References\@mkboth
 {REFERENCES}{REFERENCES}}\list
 {[\arabic{enumi}]}{\leftmargin 1em\labelwidth\z@\labelsep\z@\itemindent -1em
 \parsep -0.7ex
 \usecounter{enumi}}
 \def\newblock{\hskip .11em plus .33em minus -.07em}
 \sloppy\clubpenalty4000\widowpenalty4000
 \sfcode`\.=1000\relax}
\makeatother

\begin{document}

\title{Far-UV Line Strengths in Elliptical Galaxies}

\author{Henry C. Ferguson}
\affil{Space Telescope Science Institute}

\author{Thomas M. Brown, Arthur F. Davidsen}
\affil{Department of Physics and Astronomy, The Johns Hopkins University}




\begin{abstract}
Much of the far-UV emission from elliptical galaxies is thought to arise from
extreme horizontal branch stars and related objects. Only about 10\% of the
stellar population needs to evolve through this phase even in galaxies with the
strongest UV upturn. However it is not yet clear if this population represents
the extreme low-metallicity or high-metallicty tail of the distribution, or
rather arises from the overall population through some metallicity-insensitive
mechanism that causes increased mass loss in a small fraction of RGB stars. We
investigate the utility of far-UV line strengths for deciding between these
possiblities.  Complications include the fact that the line strengths reflect
both the temperature distribution and the metallicity distribution of the
stars, that there may be abundance anomalies introduced on the RGB, and that
metals are likely to be redistributed by gravitational settling 
and radiative diffusion in
the atmospheres of hot high-gravity stars. Line-strength measurements from
Astro-2 HUT spectra are considered in this context.
\end{abstract}


\keywords{}


\section{Introduction}
Giant elliptical galaxies show a large variation in the ratios of
their far-UV to optical fluxes. Shortward of 2000{\AA}, most ellipticals
have spectra that rise in $f_\lambda$ toward shorter wavelengths.
This hot component has been known since the early days of space astronomy
\cite{CW79}, but it is only within the last two years that observations
and theory seem to be converging on a consensus that the dominant
component in UV bright galaxies is extreme horizontal branch 
(EHB) stars and their evolutionary progeny \cite{FD93,DOR95,BCF94,BFD95}. 
This conclusion stems from the rather cool
temperature (25000 K) derived for the dominant component in NGC1399
\cite{Ferg91L}, and from computations that indicate that EHB stars
can provide enough far-UV photons over their lifetimes to produce the
elliptical galaxy fluxes, while other candidates such as PAGB stars cannot
\cite{GR90,DOR95}.

While it seems clear that EHB stars provide the far-UV flux, it is
not at all clear how they got there. The general trend observed
for globular clusters is that the horizontal branch (HB) becomes redder
with increasing metallicity. Elliptical galaxies are even 
more metal rich than Galactic globular clusters; they must somehow
be able to buck the trend. The HB morphology depends 
on age, metallicity, helium abundance, and the amount of mass loss
on the red giant branch. The helium-burning core in HB
stars has a mass (0.5 $M_\odot$) that is nearly independent of these
parameters. The position of stars along the HB
thus depends on the envelope mass, which in turn depends on the main-sequence
mass and the amount of mass lost during the RGB phase. 
EHB stars (those
with $T_{eff} > 20000$~K) have envelope masses less than 0.05$M_\odot$. 
Hence they must arise from stars that have lost nearly all the mass it
was possible for them to lose and still ignite helium in their cores.

There are several plausible ways to produce a minority population of
EHB stars in elliptical galaxies. 

First, the giant elliptical galaxies
may in general be {\it older} than the galactic globular clusters.
This argument is a natural extension of the interpretation of the
second parameter effect in globular clusters as being due to
variations in age 
\cite{Lee94,PL95}. 
The EHB stars in this model represent
the extreme metal-poor tail of the metallicity distribution, and are
2-4 Gyr older than the most-metal poor globular clusters. They show
up in giant elliptical galaxies, which are on average metal rich, because
these galaxies formed first, and hence have the oldest stars. 

Second, elliptical galaxies may have high helium abundance $Y$
\cite{GR90,BCF94,YADO95}. 
At fixed age and metallicity, the main-sequence
lifetime decreases with increasing $Y$. Observations of nearby 
star-forming galaxies and the galactic bulge hint at a rather steep
relation ($\Delta Y / \Delta Z > 2$)
between helium abundance and metallicity \cite{Pagel89p201,Renzini94}.
If this is the case, then old metal-rich populations may have EHB stars,
even with standard RGB mass-loss rates. The required ages ($>7$ Gyr)
are not as extreme as in the Lee model. In these models, the EHB stars
arise from the extreme high-metallicity tail of the abundance distribution.

Third, some other process may act to increase mass loss in a small
fraction of the population. For example, EHB stars could arise from
stars in close binary systems that have shed their envelopes during
interactions with their companions, our they could
arise only from stars with high rotation rates. Such mechanisms could
produce EHB populations from anywhere in the abundance distribution,
but might be enhanced in giant ellipticals through some secondary
effect (for example binary fraction might somehow depend on galaxy 
metallicity or velocity dispersion). In this case, the EHB stars may come closer to 
reflecting the mean metallicity of the stellar population. 

To distinguish between these possibilities, it is important to 
try to get some direct measure of the metallicities of the far-UV
emitting population, as suggested for example by \citeN{PL95} and
\citeN{YADO95}. It is now possible to attempt this with new observations
from the Hopkins Ultraviolet Telescope (HUT), obtained March 1995 during
the Astro-2 mission. In the rest of this contribution we summarize
our preliminary attempts to do this, and outline some arguments why such
efforts may in the end yield ambiguous results.

\section{Observations}

Elliptical galaxies were among the prime targets for the Hopkins
Ultraviolet Telescope (HUT) on both the Astro-1 and Astro-2 missions.
The HUT spectra cover the wavelength range from 1820{\AA} to the 
Lyman limit at a resolution of about 3{\AA}. \citeN{BFD95}
present spectra for six galaxies observed on the Astro-2 mission.
Some absorption line features are clearly visible in the spectra
of individual galaxies, including strong Ly$\beta$ and Ly$\gamma$, and
clearly visible CIII] 1175. To improve the S/N to the point where
weaker features are measurable, we have summed together the two
UV-bright galaxies: NGC1399, from Astro-1, and NGC4649 (M60), 
from Astro-2.
For comparison, we have summed together the UV-weaker galaxies
NGC3379, NGC4472 (M49), and NGC3115. The Astro-1 M31 spectrum, although
of reasonably high S/N, is not included in this comparison
because (1) it is contaminated by interstellar lines due to its
low galactic latitude and low redshift, and (2) its far-UV
spectral energy distribution is significantly different from that of
other galaxies with similar 1550-V colors. 

The Astro-2 observing program also included several Galactic sdB
stars from the survey by \citeN{SBKL94}.
These stars are local examples of the stars that might be
producing the far-UV emission in elliptical galaxies. 
(Their existence in the disk of our galaxy is at least
as puzzling as the existence of EHB stars in elliptical galaxies.) The
sdB star observations were included to provide a sensitive test of
model atmospheres in the temperature and gravity range relevant to
elliptical galaxies.  The sdB star PG1710+490, with $T_{\rm eff} =
29900$~K by itself provides a fairly direct comparison to a star at roughly
solar metallicity.

Line strengths are sensitive to both metallicity and temperature.
To try to separate the two effects, we have used the \citeN{Hubeny88}
TLUSTY and SYNSPEC codes to construct line-blanketed spectra for
a grid of parameters (T$_{eff}$, log(g), and [Fe/H]) which cover the range
relevant to the Dorman EHB models.
For T$_{eff}\leq~20,000$~K,
we used the Kurucz (1993) atmospheric structure as SYNSPEC input.  For the ranges
20,000~K$<$T$_{eff}<~45,000$~K and T$_{eff}\geq~45,000$~K, we
generated LTE and NLTE atmospheres of pure H + He, and used them as SYNSPEC
input.  SYNSPEC then generates the emergent spectrum using line opacities
from those elements with atomic number Z$\leq~28$.  The grid spacing
in gravity is 0.25 and 0.5, respectively, for the Hubeny and Kurucz atmospheres,
while the irregular spacing in T$_{eff}$ varies from 2000~K to 10000~K,
as one moves from cooler to hotter grid temperatures.  Although the models
are clearly only an approximation to real stellar atmospheres, the purpose
here is to identify useful wavelength ranges for testing for metal
abundance, and to see qualitativley how the line strengths vary with abundance.
The density of lines in the
spectra shortward of 1200{\AA} is so great that each HUT resolution
element contains hundreds of unresolved lines. Thus, measurements
of the line strengths do not provide an unambiguous measure of which
elements are varying, just an indication of whether the abundances are
much above or below the solar value.

\begin{figure}
\plotone{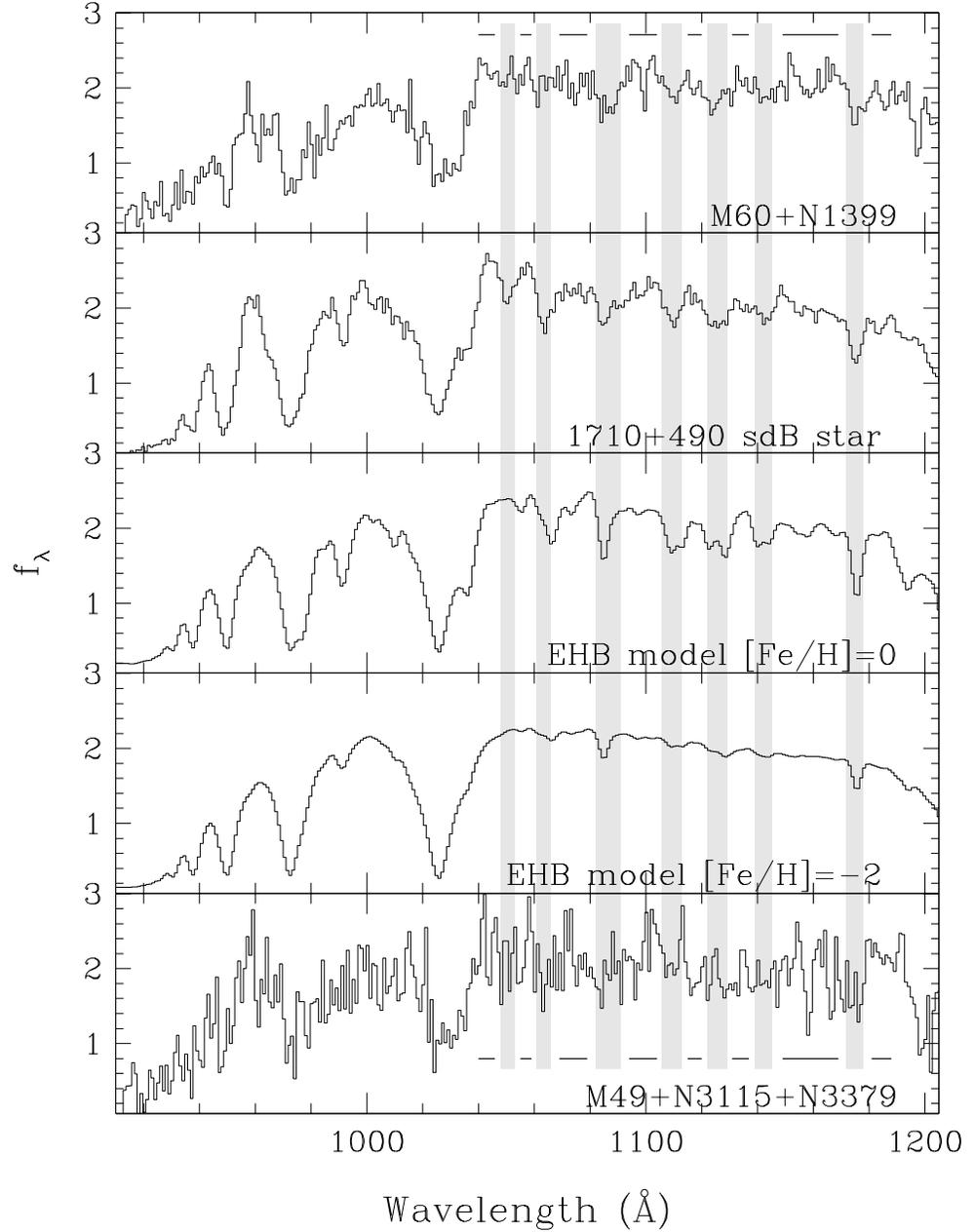}
\caption{Spectra from Lyman $\alpha$ to the Lyman limit. The top
and bottom panels show sums of elliptical galaxy spectra. The ones
in the top panel are the strongest UV emitters; the ones in the bottom
pannel are somewhat weaker. The middle three panels show comparison
spectra (described in the text). The shaded regions are used as the line
regions for the spectral index measured in Table 1. The continuum regions
are shown as horizontal lines.}
\end{figure}

In Figure 1 the summed HUT spectra are compared to PG1710+490, and to two
models constructed using the \shortciteN{DRO93} EHB evolutionary
tracks. While the evolutionary track is appropriate for abundances
$Y_{\rm ZAMS}=0.34$, $Z_{\rm ZAMS}=0.04$, for the synthetic spectra shown in Fig. 1, we have
used stellar atmospheres at solar abundance and $0.01 Z_\odot$ to
illustrate the effect of changing the atmospheric opacities while
holding the underlying temperature and luminosity distribution fixed.
The galaxy spectra
show some features in common with the sdB star, and are missing others
(most notably the strong nitrogen feature at 990{\AA}). The absorption
features appear on average slightly weaker than the sdB star or the 
solar-metallicity EHB star model, but the
galaxy spectra have more structure
than expected from a population with $\rm [Fe/H] = -2$. The Lyman
series lines are not as deep in the elliptical galaxies as in the
sdB star or the models. 
A 10-20\% flux contribution from PAGB stars, which have weak Lyman series
and metal lines, would be sufficient to bring the Lyman series lines
into agreement and to dilute the metal-line features. Thus
even if the EHB stars themselves are a metal-rich population, we might
expect to see somewhat weaker lines than in individual Galactic subdwarfs.
However, our synthetic spectra for the sdB star have Lyman series lines
that are stronger than observed, so we are not yet confident that 
the Lyman series line strengths can be used
to set quantitative limits on the PAGB contribution.

\begin{table}
\begin{center}
{\sc Table 1. Line Strengths}

\begin{tabular}{l c c c c}  
\hline\hline
Source & 1550-V & $I$ & C{\thinspace}IV & Si{\thinspace}IV \\ 
& (mag) & (mag) & E.W. {\AA} & E.W. {\AA} \\ \hline
N4649 & 2.24 & $0.13 \pm 0.07$ & $3.0 \pm 3.3$ & $2.2 \pm 4.0$ \\
N1399 & 2.05 & $0.04 \pm 0.09$ & $4.8 \pm 6.3$ & $2.5 \pm 6.3$ \\
N4649+N1399 & 2.14 & $0.11 \pm 0.06$ & $3.4 \pm 3.0$ & $2.2 \pm 3.4$ \\
N3379+N3315+N4472 & 3.64 & $0.09 \pm 0.11$& $1.6 \pm 3.8$ & $2.2 \pm 5.7$ \\
PG1710+490 & --- & $0.05 \pm 0.00$ & $2.5 \pm 0.0$ & $1.9 \pm 0.0$ \\
EHB model $Z_\odot$ & --- & $0.14 \pm 0.00$ & $1.9 \pm 0.0$ & $-0.8 \pm 0.0$ \\
EHB model $0.01 Z_\odot$ & --- & $0.01 \pm 0.00$ & $0.5 \pm 0.0$ & $-2.5 \pm 0.0$ \\
\noalign{\vskip -8pt}
\\
\hline\hline
\multicolumn{4}{l}{1550-V colors are taken from Burstein et al. 1988} \\
\multicolumn{4}{l}{$I_1$ is expressed as a magnitude. Bandpasses are shown in Fig. 1} \\
\multicolumn{4}{l}{Si{\thinspace}IV line: $\lambda 1385-1440$, continuum $\lambda 1360-1380, 1445-1500$} \\
\multicolumn{4}{l}{C{\thinspace}IV line: $\lambda 1540-1565$, continuum $\lambda 1480-1530, 1570-1620$} \\
\end{tabular}
\end{center}
\end{table}

The comparison of line strengths
is quantified in Table 1. At the low S/N of the HUT spectra, it is
difficult to measure individual lines. To improve the statistics, we
have concocted a composite metal line index based on the features seen
in the sdB star PG1710+490. The line and continuum regions are shown
in Fig. 1. The absorption features are due primarily to Si, C, and N.
Following standard practice, we express the result as a magnitude
($I = -2.5 \log f_\lambda (line) / f_\lambda(continuum)$).
For comparison to model predictions \cite{PL95,YADO95},
we also show Si{\thinspace}IV and C{\thinspace}IV equivalent widths. For
this purpose we have also used an HST FOS spectrum of NGC1399 to improve
the signal to noise ratio. The results favor roughly solar metallicity, but
metallities as low as $\rm [Fe/H] = -2$ or as high as $\rm [Fe/H] = 1$
are still allowable within the $3 \sigma$ uncertainties.

There is an extensive literature on abundances in the atmospheres of galactic
sdB stars (e.g. \citeNP{MBWF85,LWF87,BWMF88}).
These stars are observed to be strongly depleted in helium. 
For temperatures lower than $T_{\rm eff} = 35000$ K, abundances of 
C, N, and Si are close to solar. At higher temperatures, C and N 
are slightly depleted and Si is 
depleted by a factor of $10^4$. The line strengths in these non-convective,
high-gravity atmospheres
are governed primarily by a competition between downward diffusion
and radiative levitation, rather than by the intrinsic elemental 
abundances. In the absence of radiative forces, metals would quickly
diffuse out of the surface layers of EHB star atmospheres and the 
spectra would be extremely weak-lined. However, at EHB temperatures
radiation pressure is sufficient to balance the gravitational force.
Metals settle to an equilibrium depth where the lines are just 
saturated. If they settle further, the lines become unsaturated and
the radiation pressure increases, while if they move higher in 
the atmosphere the lines become strongly saturated and the radiation
pressure is no longer efficient. This equilibrium state can be upset
by winds, possibly explaining the Si deficiency at high temperatures
\cite{BWMF88}. These considerations suggest that interpretation of
the far-UV line-strength measurements may not be as straightforward
as originally suggested, and may require detailed understanding
of the atmospheres of hot subdwarfs.

\acknowledgments

We are grateful to Ben Dorman and Ivan Hubeny for their
valuable contributions to this work, and to the astronaut
and ground-support crew of the Astro-2 mission for 
making the observations possible. The HST portion
of this project was supported through grant GO-3647; the 
HUT portion by NASA contract NAS5-27000 to the Johns Hopkins
University. 

\bibliography{paspmnemonic,bib}
\bibliographystyle{pasp}

\end{document}